\documentclass[10pt,preprint]{aastex}

\def \kbf {{\bf k}}
\def \rbf {{\bf r}}
\def \xbf {{\bf x}}
\def \ybf {{\bf y}}
\def \zbf {{\bf z}}

\begin{document}

\title{Generation of Gaussian Density Fields}

\author{Hugo Martel}

\author{D\'epartement de physique, g\'enie physique et optique,
Universit\'e Laval, Qu\'ebec, QC, G1K 7P4, Canada}

\author{\tt Technical Report UL-CRC/CTN-RT003}

\begin{abstract}
This document describes analytical and numerical techniques for the 
generation of Gaussian density fields, which represent cosmological
density perturbations. The mathematical techniques involved in
the generation of density harmonics
in $k$-space, the filtering of the density fields, and the normalization
of the power spectrum to the measured temperature fluctuations
of the Cosmic Microwave Background, are presented in details.
These techniques are well-known amongst experts, but the current
literature lacks a formal description.
I hope that this technical report will prove useful
to new researchers moving into this field, sparing them the task of
reinventing the wheel.
\end{abstract}

\keywords{cosmology: theory --- methods: numerical}

\clearpage

\section{INTRODUCTION}

Gaussian density field play a major role in cosmology. There is now
strong evidence that the large-scale structure of the universe originated
from the growth, by gravitational instability, of primordial 
density fluctuations. Observations of the temperature fluctuations
of the Cosmic Microwave Background (CMB) indicate that these primordial
fluctuation were Gaussian.

The formation of evolution of large-scale structure in the
universe is a complex problem that requires a numerical approach. Typically,
one creates a realization of the density fluctuations at early time, and
uses a numerical algorithm to evolve this fluctuation all the way
to the present (or to some redshift of interest). Since it is impossible
to simulate the entire universe (which might very well be infinite), we
normally assume that the universe is periodic at large scales.
We can then divide the universe into identical cubes of volume
$V_{\rm box}=L_{\rm box}^3$, and we only need to simulate one cube. This
approximation is valid as long as the box size $L_{\rm box}$ is much larger
than any existing large-scale structure in the universe. We can rephrase
this by saying that the box must contain a ``fair'' sample of the universe.

The density field can be represented in two different ways. 
In the {\it Eulerian Representation\/}, the box is divided into
$N\times N\times N$ cells, and the density contrast $\delta$ is calculated
at the center of each cell. In the {\it Lagrangian Representation\/}, 
$N_p\times N_p\times N_p$ equal-mass particles are laid down on a cubic
grid inside the box, and are then displaced in order to represent the density
fluctuation. The choice of representation depends on the particular algorithm
that will be used to evolve the system from these initial conditions.

\subsection{The Power Spectrum}

A Gaussian density field can be represented as a superposition of plane waves
of wavevectors $\kbf$ and complex amplitudes $\delta_{\kbf}^{\rm cont}$,
where the superscript ``cont'' stands for {\it continuous},
indicating that all values of $\kbf$ are allowed.  
The amplitudes are related to the power spectrum
$P(k)$ by
\begin{equation}
\label{deltap}
|\delta_{\kbf}^{\rm cont}|^2\propto P(k)\,,
\end{equation}

\noindent where $k=|\kbf|$.
However, there is a lot of confusion in the literature about the
constant of proportionality between $|\delta_{\kbf}^{\rm cont}|^2$ and
$P(k)$, and even the {\it units\/} of $\delta_{\kbf}^{\rm cont}$ can
vary from one author to another. We will clarify this issue in \S2.

Note: Equation~(\ref{deltap}) is a convenient simplification.
As we will discuss later, in a truly Gaussian random field,
the amplitudes $\delta_{\kbf}$ are determined only in a statistical sense.
Their real and imaginary parts are separately given by a Gaussian 
distribution whose variance is proportional to $P(k)$. However, using
equation~(\ref{deltap}) greatly simplifies the derivation presented
in \S2, without affecting the results.

\section{THE AMPLITUDES OF THE DENSITY MODES}

We assume that the universe is periodic over a comoving cubic volume
$V_{\rm box}=L_{\rm box}^3$. The density contrast $\delta$ can be
decomposed into a sum of plane waves.
\begin{equation}
\label{waves}
\delta(\rbf)={1\over N^3}\sum_\kbf\delta_\kbf^{\rm disc}e^{-i\kbf\cdot\rbf}\,,
\end{equation}

\noindent where $\rbf$ is the comoving position. The wavevectors
$\kbf$ are given by
\begin{equation}
\label{wavevect1}
\kbf=(l,m,n)k_0\,,\qquad l,m,n=-\infty,\ldots,-1,0,1,\ldots,\infty\,.
\end{equation}

\noindent where the fundamental wavenumber is
\begin{equation}
k_0={2\pi\over L_{\rm box}}\,.
\end{equation}

\noindent
The superscript ``${\rm disc}$'' indicates that the amplitudes 
$\delta_\kbf^{\rm disc}$ form a discrete spectrum, that is, they are defined
for particular, discrete values of $\kbf$.\footnote{Other values of $\kbf$
would not satisfy the periodic boundary conditions}
The factor $1/N^3$ is not necessary at this point, and could be
absorbed into the definition of $\delta_\kbf^{\rm disc}$. We introduce
it to make the notation consistent with \S3.
To make $\delta(\rbf)$ real, the coefficients
$\delta_\kbf^{\rm disc}$ must satisfy the {\it reality condition}:
\begin{equation}
\label{reality}
\delta_{-\kbf}^{\rm disc}=(\delta_\kbf^{\rm disc})^*\,.
\end{equation}

Our first challenge is to relate the discrete sum in equation~(\ref{waves})
to the continuous sum of modes present in the real universe, and to express
the amplitudes $\delta_\kbf^{\rm disc}$ in terms of the power spectrum $P(k)$.
To do so, we will
consider the rms fluctuation of the density at a certain scale $R$, and
match the expressions obtained in the discrete and continuous limits.

\subsection{The Discrete Limit}

Consider a sphere of radius $R$ centered at $\rbf_0$. 
The mass inside that sphere is given by
\begin{equation}
M(\rbf_0)=\int_{{\rm sph}(\rbf_0)}\bar\rho\left[1+\delta(\rbf)\right]d^3r
=\bar\rho V_{\rm sph}+{\bar\rho\over N^3}
\int_{{\rm sph}(\rbf_0)}d^3r\sum_\kbf\delta_\kbf^{\rm disc}e^{-i\kbf\cdot\rbf}
\,,
\end{equation}

\noindent where $\bar\rho$ is the average density,
$V_{\rm sph}=4\pi R^3/3$ 
is the volume of the sphere, and the integral is computed
over that volume. The relative mass excess in the sphere is given by
\begin{equation}
\label{dmm}
{\Delta M\over M}(\rbf_0)={M(\rbf_0)-\bar\rho V_{\rm sph}
\over\bar\rho V_{\rm sph}}=
{1\over N^3V_{\rm sph}}
\int_{{\rm sph}(\rbf_0)}d^3r\sum_\kbf\delta_\kbf^{\rm disc}e^{-i\kbf\cdot\rbf}
\,.
\end{equation}

\noindent We introduce the following change of variables,
\begin{equation}
\rbf=\rbf_0+\ybf\,.
\end{equation}

\noindent In $\ybf$-space, the sphere is now located at the origin, and 
equation~(\ref{dmm}) becomes
\begin{equation}
{\Delta M\over M}(\rbf_0)={1\over N^3V_{\rm sph}}
\int_{{\rm sph}(0)}d^3y\sum_\kbf\delta_\kbf^{\rm disc}e^{-i\kbf\cdot\rbf_0}
e^{-i\kbf\cdot\ybf}\,.
\end{equation}

\noindent We now square this expression, and get
\begin{equation}
\biggl({\Delta M\over M}\biggr)^2(\rbf_0)={9\over16\pi^2R^6N^6}
\Biggl[\int_{\rm sph(0)}d^3y\sum_\kbf\delta_\kbf^{\rm disc}
e^{-i\kbf\cdot\rbf_0}e^{-i\kbf\cdot\ybf}\Biggr]
\Biggl[\int_{\rm sph(0)}d^3z\sum_{\kbf'}\delta_{\kbf'}^{\rm disc}
e^{-i\kbf'\cdot\rbf_0}e^{-i\kbf'\cdot\zbf}\Biggr]\,.
\end{equation}

\noindent The variance of the density
contrast at scale $R$ is obtained by averaging the above expression over 
all possible locations $\rbf_0$ of the sphere inside the computational box,
\begin{eqnarray}
\sigma_R^2&\equiv&
\Biggl\langle\biggl({\Delta M\over M}\biggr)^2\Biggr\rangle_{V_{\rm box}}
={1\over V_{\rm box}}\int_{V_{\rm box}}d^3r_0
\biggl({\Delta M\over M}\biggr)^2(\rbf_0)\nonumber\\
\label{sx2}
&=&{1\over V_{\rm box}}{9\over16\pi^2R^6N^6}\int_{V_{\rm box}}d^3r_0
\int_{\rm sph(0)}d^3y\int_{\rm sph(0)}d^3z\sum_\kbf\sum_{\kbf'}
\delta_\kbf^{\rm disc}\delta_{\kbf'}^{\rm disc}
e^{-i\kbf\cdot\ybf}e^{-i\kbf'\cdot\zbf}
e^{-i(\kbf+\kbf')\cdot\rbf_0}\,.
\end{eqnarray}

\noindent The integral over $V_{\rm box}$ reduces to
\begin{equation}
\label{dkk}
\int_{V_{\rm box}}d^3r_0e^{-i(\kbf+\kbf')\cdot\rbf_0}=
V_{\rm box}\delta_{\kbf,-\kbf'}\,.
\end{equation}

\noindent We substitute this expression in equation~(\ref{sx2}), and use the
Kronecker~$\delta$ to eliminate the summation over $\kbf'$. 
Equation~(\ref{sx2})
reduces to
\begin{equation}
\label{sx2b}
\sigma_R^2={9\over16\pi^2R^6N^6}\sum_\kbf|\delta_\kbf^{\rm disc}|^2
\Biggl[\int_{\rm sph(0)}d^3y\,e^{-i\kbf\cdot\ybf}\Biggr]^2\,,
\end{equation}

\noindent where we used equation~(\ref{reality}) to get
$\delta_{-\kbf}^{\rm disc}\delta_\kbf^{\rm disc}
=(\delta_\kbf^{\rm disc})^*\delta_\kbf^{\rm disc}
=|\delta_\kbf^{\rm disc}|^2$.
The remaining integral can be evaluated easily
(see Appendix A). Equation~(\ref{sx2b})
reduces to
\begin{equation}
\label{sx2c}
\sigma_R^2={1\over N^6}\sum_\kbf|\delta_\kbf^{\rm disc}|^2W^2(kR)\,,
\end{equation}

\noindent where
\begin{equation}
W(u)\equiv{3\over u^3}(\sin u-u\cos u)\,.
\end{equation}

\subsection{The Continuous Limit}

The real universe is of course not periodic, in which case all values
of $\kbf$ are allowed. To convert the expressions derived in \S2.1 from the
discrete limit to the continuous one, consider any function $f_\kbf$ that
is summed over all allowed values of $\kbf$. In the discrete limit, we have
\begin{equation}
\label{sumfkdisc}
\sum_\kbf f_\kbf^{\rm disc}=\sum_{\rm all\>V.E.} f_\kbf^{\rm disc}
={1\over k_0^3}\sum_{\rm all\>V.E.} f_\kbf^{\rm disc}k_0^3
={V_{\rm box}\over(2\pi)^3}\sum_{\rm all\>V.E.} f_\kbf^{\rm disc}
\int_{\rm V.E.}d^3k\,,
\end{equation}

\noindent where ``V.E.'' represent a {\it volume element\/} in $\kbf$-space,
which is a cube of volume $k_0^3$ centered around an allowed value of $\kbf$
(with $k_0=2\pi/L_{\rm box}$). Assuming that the
function $f_\kbf^{\rm disc}$ does not vary significantly 
over one volume element, we can pull
it inside the integral,
\begin{equation}
\label{sumfkdisc2}
\sum_\kbf f_\kbf^{\rm disc}\approx{V_{\rm box}\over(2\pi)^3}
\sum_{\rm all\>V.E.}\int_{\rm V.E.}f_\kbf^{\rm disc}d^3k\,.
\end{equation}

\noindent Of course, the effect of integrating over the volume element, and
then summing over all volume elements, is to effectively integrate
over all $k$-space, so equation~(\ref{sumfkdisc2}) reduces to
\begin{equation}
\sum_\kbf f_\kbf^{\rm disc}\approx{V_{\rm box}\over(2\pi)^3}
\int f_\kbf^{\rm disc}d^3k\,.
\end{equation}

\noindent We can rewrite this expression as
\begin{equation}
\sum_\kbf f_\kbf^{\rm disc}\approx\int f_\kbf^{\rm cont}d^3k\,,
\end{equation}

\noindent where the continuous and discrete functions are related by
\begin{equation}
f_\kbf^{\rm cont}={V_{\rm box}\over(2\pi)^3}f_\kbf^{\rm disc}\,.
\end{equation}

\noindent Using these formulae, we can rewrite equation~(\ref{waves}) as
\begin{equation}
\delta(\rbf)={1\over N^3}
\int d^3k\,\delta_\kbf^{\rm cont}e^{-i\kbf\cdot\rbf}\,,
\end{equation}

\noindent where
\begin{equation}
\label{dk}
\delta_\kbf^{\rm cont}={V_{\rm box}\over(2\pi)^3}\delta_\kbf^{\rm disc}\,.
\end{equation}

\noindent 
Let us now convert equation~(\ref{sx2c}) 
into an integral, as we did for equation~(\ref{sumfkdisc}).
We get
\begin{equation}
\label{sx2d}
\sigma_R^2={V_{\rm box}\over(2\pi)^3N^6}\int d^3k
|\delta_\kbf^{\rm disc}|^2W^2(kR)\,.
\end{equation}

\noindent We substitute equation~(\ref{dk}) 
into equation~(\ref{sx2d}), and get
\begin{equation}
\label{sx2e}
\sigma_R^2={(2\pi)^3\over V_{\rm box}N^6}\int d^3k
|\delta_\kbf^{\rm cont}|^2W^2(kR)\,.
\end{equation}

\noindent We then need to related $\sigma_R^2$ to the power
spectrum $P(k)$. The relation is given by \cite{bw97} as
\begin{equation}
\label{sx2e2}
\sigma_R^2={1\over2\pi^2}\int_0^\infty dk\,k^2P(k)W^2(kR)\,.
\end{equation}

\noindent This relation is obtained by performing an integration
over angles, using the fact that $P(\kbf)$ is a function of $k=|\kbf|$
only. We can ``undo'' this integration, simply by dividing
equation~(\ref{sx2e2}) by $4\pi$.  We get

\begin{equation}
\label{sx2f}
\sigma_R^2={1\over(2\pi)^3}\int d^3k\,P(k)W^2(kR)\,.
\end{equation}

\noindent Comparing equations~(\ref{sx2d}), (\ref{sx2e}), 
and~(\ref{sx2f}), we get
\begin{equation}
\label{pdelta}
P(k)={V_{\rm box}\over N^6}|\delta_\kbf^{\rm disc}|^2=
{(2\pi)^6\over V_{\rm box}N^6}|\delta_\kbf^{\rm cont}|^2\,.
\end{equation}

\noindent This gives us the relation between $P(k)$,
$\delta_\kbf^{\rm disc}$, and $\delta_\kbf^{\rm cont}$.
Notice the $P(k)$ is neither the square of 
$\delta_\kbf^{\rm disc}$ nor the square of $\delta_\kbf^{\rm cont}$.
Both $P(k)$ and $\delta_\kbf^{\rm cont}$ have dimensions of a volume
while $\delta_\kbf^{\rm disc}$ is dimensionless. 

Equations~(\ref{sx2e2}) and (\ref{sx2f}) define the normalization of the
power spectrum. When using any power spectrum obtained from the literature,
{\it it is essential to check that these relations are satisfied}. Variations
by factors of $2\pi$ between different papers are quite common.

\section{DIRECT CALCULATION OF THE DENSITY HARMONICS}

Using the formalism described in \S2,
we now want to compute the
density harmonics $\hat\delta_\kbf$. We first lay down inside the
computational volume $V_{\rm box}$ a regular 
cubic grid of size $N\times N\times N$, with grid spacing 
$\Delta=L_{\rm box}/N$.
The coordinates ${\bf r}$ of the grid points are given by
\begin{equation}
\rbf=(\alpha,\beta,\gamma)\Delta\,,\qquad\alpha,\beta,\gamma=0,1,\ldots,
N-1\,.
\end{equation}

\noindent The presence of a grid results in a discretization of space,
which in turns modifies the structure of the $k$-space. 
In equation~(\ref{wavevect1}), the values of $\kbf$ form an infinite
cubic grid in $k$-space, since the indicies $l$, $m$, $n$ can take any 
integer value from
$-\infty$ to $+\infty$. However, the discretization of space limits
the number of modes. Consider a mode with wavenumber
\begin{equation}
\kbf'=\kbf+(uN,vN,wN)k_0\,,
\end{equation}

\noindent where $u$, $v$, and $w$ are integers. The exponential in 
equation~(\ref{waves}) becomes
\begin{eqnarray}
e^{-i\kbf'\cdot\rbf}
&=&e^{-i\kbf\cdot\rbf}e^{-i[uN,vN,wN]\cdot[\alpha,\beta,\gamma]k_0\Delta}
=e^{-i\kbf\cdot\rbf}e^{-i[uN,vN,wN]\cdot[\alpha,\beta,\gamma](2\pi/N)}
\nonumber\\
&=&e^{-i\kbf\cdot\rbf}e^{-2\pi i[u,v,w]\cdot[\alpha,\beta,\gamma]}
=e^{-i\kbf\cdot\rbf}\,.
\end{eqnarray}

\noindent Hence, the modes $\kbf'$ and $\kbf$ are {\it inseparable}. They
represent a plane wave with the same effective wavenumber. Consequently, we
will consider a finite $k$-space, in which the values $l$, $m$, $n$ do not
run from $-\infty$ to $+\infty$, but instead are limited to the range 0
to $N-1$. Hence, in $k$-space, the density harmonics $\hat\delta({\bf k})$
are also defined on a regular cubic grid of size $N\times N\times N$ .
The wavevectors are given by
\begin{equation}
\label{kallowed}
\kbf=(l,m,n)k_0\,,\qquad l,m,n=0,1,\ldots,N-1\,.
\end{equation}

\noindent In doing so, we are
simply neglecting high-frequency modes. This is justified since the
discreteness of the grid in $r$-space prevents us from resolving any
structure that these modes represent. Hence, by using a grid in
$r$-space, we are effectively performing
a filtering of the density fluctuation at the scale of $\Delta$, and such
filtering eliminates high-frequency modes.

The Fourier transform and inverse Fourier transform are given
respectively by
\begin{eqnarray}
\label{fft}
\hat\delta(\kbf)&=&\sum_\rbf\delta(\rbf)e^{i\kbf\cdot\rbf}\,,\\
\label{fft2}
\delta(\rbf)&=&{1\over N^3}
\sum_\kbf\hat\delta(\kbf)e^{-i\kbf\cdot\rbf}\,.
\end{eqnarray}

\noindent Notice that equation~(\ref{fft2}) is the same as
equation~(\ref{waves}), with a slight change of notation:
$\delta_\kbf^{\rm disc}\rightarrow\hat\delta(\kbf)$.
We introduce the notation
\begin{eqnarray}
\hat\delta(\kbf)&=&\hat\delta_{lmn}\,,\\
\delta(\rbf)&=&\delta_{\alpha\beta\gamma}\,.
\end{eqnarray}

\noindent Equation~(\ref{fft}) becomes
\begin{eqnarray}
\hat\delta_{lmn}&=&\sum_{\alpha,\beta,\gamma=0}^{N-1}\delta_{\alpha\beta\gamma}
e^{i(2\pi/\Delta N)\Delta[l,m,n]\cdot[\alpha,\beta,\gamma]}
=\sum_{\alpha,\beta,\gamma=0}^{N-1}\delta_{\alpha\beta\gamma}
e^{2\pi il\alpha/N}e^{2\pi im\beta/N}e^{2\pi in\gamma/N} \nonumber \\
\label{mode1}
&=&\sum_{\alpha,\beta,\gamma=0}^{N-1}
\left(\cos{2\pi l\alpha\over N}+i\sin{2\pi l\alpha\over N}\right)
\left(\cos{2\pi m\beta\over N}+i\sin{2\pi m\beta\over N}\right)
\left(\cos{2\pi n\gamma\over N}+i\sin{2\pi n\gamma\over N}\right)\,.
\end{eqnarray}

\noindent After expansion, this expression becomes
\begin{equation}
\label{dlmn}
\hat\delta_{lmn}=
(\hat\delta_{eee}+\hat\delta_{eoo}+\hat\delta_{oeo}+\hat\delta_{ooe})
+i(\hat\delta_{eeo}+\hat\delta_{eoe}+\hat\delta_{oee}+\hat\delta_{ooo})\,,
\end{equation}

\noindent where we define
\begin{eqnarray}
\label{deltaeee}
\hat\delta_{eee}&\equiv&\sum_{\alpha,\beta,\gamma=0}^{N-1}
                       \delta_{\alpha\beta\gamma}
                       \cos{2\pi l\alpha\over N}
                       \cos{2\pi m\beta\over N}
                       \cos{2\pi n\gamma\over N}\,,\\
\hat\delta_{eeo}&\equiv&\sum_{\alpha,\beta,\gamma=0}^{N-1}
                       \delta_{\alpha\beta\gamma}
                       \cos{2\pi l\alpha\over N}
                       \cos{2\pi m\beta\over N}
                       \sin{2\pi n\gamma\over N}\,,\\
\hat\delta_{eoe}&\equiv&\sum_{\alpha,\beta,\gamma=0}^{N-1}
                       \delta_{\alpha\beta\gamma}
                       \cos{2\pi l\alpha\over N}
                       \sin{2\pi m\beta\over N}
                       \cos{2\pi n\gamma\over N}\,,\\
\hat\delta_{eoo}&\equiv&-\sum_{\alpha,\beta,\gamma=0}^{N-1}
                       \delta_{\alpha\beta\gamma}
                       \cos{2\pi l\alpha\over N}
                       \sin{2\pi m\beta\over N}
                       \sin{2\pi n\gamma\over N}\,,\\
\hat\delta_{oee}&\equiv&\sum_{\alpha,\beta,\gamma=0}^{N-1}
                       \delta_{\alpha\beta\gamma}
                       \sin{2\pi l\alpha\over N}
                       \cos{2\pi m\beta\over N}
                       \cos{2\pi n\gamma\over N}\,,\\
\hat\delta_{oeo}&\equiv&-\sum_{\alpha,\beta,\gamma=0}^{N-1}
                       \delta_{\alpha\beta\gamma}
                       \sin{2\pi l\alpha\over N}
                       \cos{2\pi m\beta\over N}
                       \sin{2\pi n\gamma\over N}\,,\\
\hat\delta_{ooe}&\equiv&-\sum_{\alpha,\beta,\gamma=0}^{N-1}
                       \delta_{\alpha\beta\gamma}
                       \sin{2\pi l\alpha\over N}
                       \sin{2\pi m\beta\over N}
                       \cos{2\pi n\gamma\over N}\,,\\
\label{deltaooo}
\hat\delta_{ooo}&\equiv&-\sum_{\alpha,\beta,\gamma=0}^{N-1}
                       \delta_{\alpha\beta\gamma}
                       \sin{2\pi l\alpha\over N}
                       \sin{2\pi m\beta\over N}
                       \sin{2\pi n\gamma\over N}\,.
\end{eqnarray}

Consider now the mode $\hat\delta_{N-l,m,n}$. We replace $l$ by
$N-l$ in equation~(\ref{mode1}), and get
\begin{equation}
\label{mode2}
\hat\delta_{N-l,mn}
=\sum_{\alpha,\beta,\gamma=0}^{N-1}\delta_{\alpha\beta\gamma}
e^{2\pi i(N-l)\alpha/N}e^{2\pi im\beta/N}e^{2\pi in\gamma/N} 
=\sum_{\alpha,\beta,\gamma=0}^{N-1}\delta_{\alpha\beta\gamma}
e^{2\pi i\alpha}e^{-2\pi il\alpha/N}e^{2\pi im\beta/N}e^{2\pi in\gamma/N}\,.
\end{equation}

\noindent Since $\alpha$ is an integer, the first exponential is always
unity, and equation~(\ref{mode2}) reduces to
\begin{equation}
\label{mode2b}
\hat\delta_{N-l,mn}=\sum_{\alpha,\beta,\gamma=0}^{N-1}
\left(\cos{2\pi l\alpha\over N}-i\sin{2\pi l\alpha\over N}\right)
\left(\cos{2\pi m\beta\over N}+i\sin{2\pi m\beta\over N}\right)
\left(\cos{2\pi n\gamma\over N}+i\sin{2\pi n\gamma\over N}\right)\,.
\end{equation}

\noindent Comparing equations (\ref{mode1}) 
and (\ref{mode2b}), we see that the effect of 
replacing $l$ by $N-l$ amounts to a change of sign of the first sine function.
In equations (\ref{deltaeee})--(\ref{deltaooo}), 
that sine appears only in the $\hat\delta$'s for
which the first subscript is ``$o$.'' Hence, these $\hat\delta$'s will
change sign, and equation~(\ref{dlmn}) will become
\begin{equation}
\label{dnlmn}
\hat\delta_{N-l,mn}=
(\hat\delta_{eee}+\hat\delta_{eoo}-\hat\delta_{oeo}-\hat\delta_{ooe})
+i(\hat\delta_{eeo}+\hat\delta_{eoe}-\hat\delta_{oee}-\hat\delta_{ooo})\,,
\end{equation}

We can directly generalize to the other indicies, or combinations of them.
Replacing $m$ by $N-m$ changes the sign of the $\hat\delta$'s for
which the second subscript is ``$o$,'' and
replacing $n$ by $N-n$ changes the sign of the $\hat\delta$'s for
which the third subscript is ``$o$.'' Hence,
\begin{eqnarray}
\label{dlnmn}
\hat\delta_{l,N-m,n}&=&
(\hat\delta_{eee}-\hat\delta_{eoo}+\hat\delta_{oeo}-\hat\delta_{ooe})
+i(\hat\delta_{eeo}-\hat\delta_{eoe}+\hat\delta_{oee}-\hat\delta_{ooo})\,,\\
\label{dlmnn}
\hat\delta_{lm,N-n}&=&
(\hat\delta_{eee}-\hat\delta_{eoo}-\hat\delta_{oeo}+\hat\delta_{ooe})
+i(-\hat\delta_{eeo}+\hat\delta_{eoe}+\hat\delta_{oee}-\hat\delta_{ooo})\,,\\
\hat\delta_{N-l,N-m,n}&=&
(\hat\delta_{eee}-\hat\delta_{eoo}-\hat\delta_{oeo}+\hat\delta_{ooe})
+i(\hat\delta_{eeo}-\hat\delta_{eoe}-\hat\delta_{oee}+\hat\delta_{ooo})\,,\\
\hat\delta_{N-l,m,N-n}&=&
(\hat\delta_{eee}-\hat\delta_{eoo}+\hat\delta_{oeo}-\hat\delta_{ooe})
+i(-\hat\delta_{eeo}+\hat\delta_{eoe}-\hat\delta_{oee}+\hat\delta_{ooo})\,,\\
\label{dlnmnn}
\hat\delta_{l,N-m,N-n}&=&
(\hat\delta_{eee}+\hat\delta_{eoo}-\hat\delta_{oeo}-\hat\delta_{ooe})
+i(-\hat\delta_{eeo}-\hat\delta_{eoe}+\hat\delta_{oee}+\hat\delta_{ooo})\,,\\
\label{dnlnmnn}
\hat\delta_{N-l,N-m,N-n}&=&
(\hat\delta_{eee}+\hat\delta_{eoo}+\hat\delta_{oeo}+\hat\delta_{ooe})
+i(-\hat\delta_{eeo}-\hat\delta_{eoe}-\hat\delta_{oee}-\hat\delta_{ooo})\,.
\end{eqnarray}

\begin{figure}[t]
\vskip-0.5in
\hskip0.3in
\includegraphics[width=6.2in]{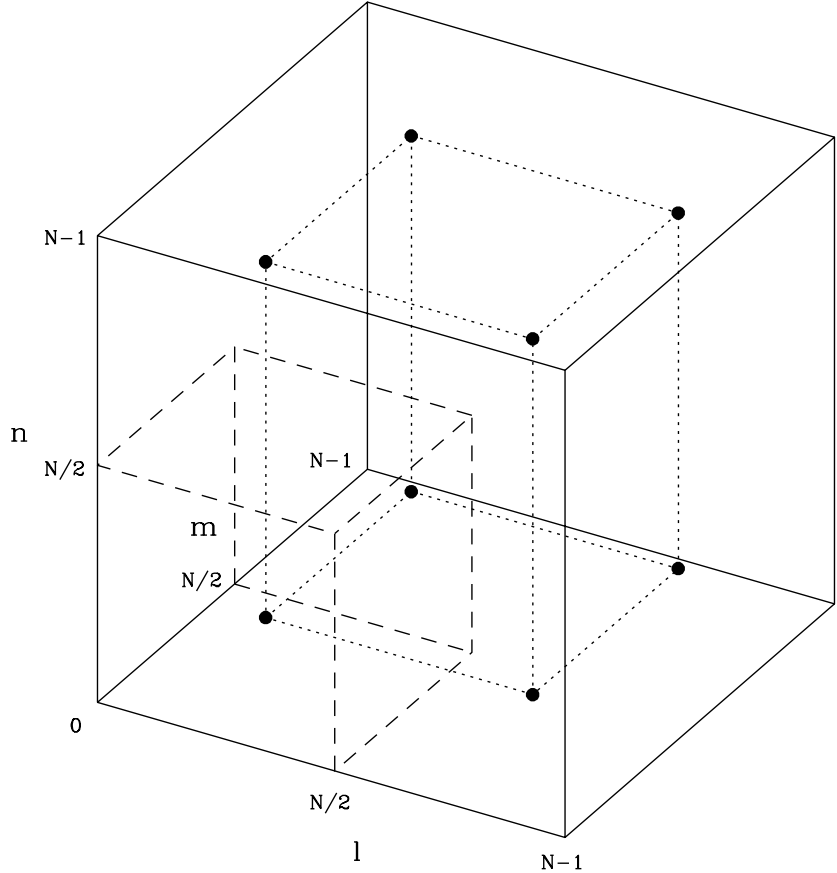}
\vskip-0.7in
\caption{Representation of the system in $k$-space. The 3 axes correspond
to the 3 indicies $l$, $m$, $n$, which run from 0 to $N-1$, with $N=64$ in
this case. The dots
indicate the particular mode $(l,m,n)=(14,15,8)$ and the seven related
modes. Notice that only one of these modes is located in the first octant,
indicated by the dashed lines.}
\label{cube1}
\end{figure}

\noindent
Hence, 8 different, but related
harmonics can be represented by various combinations
of 8 numbers: $\hat\delta_{eee}$, $\hat\delta_{eeo}$, $\hat\delta_{eoe}$, 
$\hat\delta_{oee}$, $\hat\delta_{eoo}$, $\hat\delta_{oeo}$, $\hat\delta_{ooe}$,
and $\hat\delta_{ooo}$. This implies that the Fourier transform of a
real field defined on a grid
$N\times N\times N$ can be represented by $N^3$ real numbers stored
on a similar grid, even though the Fourier transform $\hat\delta({\bf k}$)
is complex. Notice also that the 8 ``related'' modes are located, in $k$-space,
at the verticies of a rectangular box centered on the center of the $k$-space
grid ($l,m,n=N/2$). This is illustrated in Figure~\ref{cube1}.

From equations (\ref{dlmn}) and 
(\ref{dnlmn})--(\ref{dnlnmnn}), we see that related
modes form 4 pairs of complex conjugates:
\begin{eqnarray}
\label{dlmn2}
\hat\delta_{lmn}&=&\hat\delta_{N-l,N-m,N-n}^*\,,\\
\hat\delta_{lm,N-n}&=&\hat\delta_{N-l,N-m,n}^*\,,\\
\hat\delta_{l,N-m,n}&=&\hat\delta_{N-l,m,N-n}^*\,,\\
\label{dlnmnn2}
\hat\delta_{l,N-m,N-n}&=&\hat\delta_{N-l,mn}^*\,.
\end{eqnarray}

\subsection{General Case}

Consider first the modes for which the indicies $l$, $m$, and $n$ are
neither 0 nor $N/2$. In equations (\ref{dlmn2})--(\ref{dlnmnn2}), 
the amplitudes $\hat\delta$
are provided by the power spectrum. From equation~(\ref{pdelta}),
we get
\begin{equation}
\label{standard}
|\hat\delta_\kbf^{\rm disc}|=N^3\left[{P(k)\over V_{\rm box}}\right]^{1/2}\,.
\end{equation}

\noindent Equation~(\ref{standard}) provides the correct normalization
of the power spectrum. However, there are two problems with this equation.
First, it provides no mean of determining the phases of the complex
numbers $\hat\delta_\kbf^{\rm disc}$, and second, choosing random
phases would result in a field that is not Gaussian. In a truly
Gaussian field, equation~(\ref{standard}) is only valid in a statistical
sense. The correct approach is to compute the real and imaginary parts
of $\hat\delta_\kbf^{\rm disc}$ independently, by drawing them
from a Gaussian distribution,
\begin{eqnarray}
\label{real}
{\rm Re}\,\hat\delta_\kbf^{\rm disc}
&=&G_1(0,1)N^3\left[{P(k)\over2V_{\rm box}}\right]^{1/2}\,,\\
\label{imaginary}
{\rm Im}\,\hat\delta_\kbf^{\rm disc}
&=&G_2(0,1)N^3\left[{P(k)\over2V_{\rm box}}\right]^{1/2}\,,
\end{eqnarray}

\noindent where $G_1(0,1)$ and $G_2(0,1)$ are random numbers drawn from
a Gaussian distribution with mean~0 and standard deviation~1. This
ensures that the resulting field is Gaussian.

Equations~(\ref{real}) and~(\ref{imaginary}) provide us with the 
left-hand-sides of equations~(\ref{dlmn2})--(\ref{dlnmnn2}).
By equating these expressions with 
equations~(\ref{dlmn}), (\ref{dlnmn}), (\ref{dlmnn}), and (\ref{dlnmnn}),
and considering the real and imaginary parts separately, we get 8 equations,

\begin{eqnarray}
\hat\delta_{eee}+\hat\delta_{eoo}+\hat\delta_{oeo}+\hat\delta_{ooe}&=&R_1\,,\\
\hat\delta_{eeo}+\hat\delta_{eoe}+\hat\delta_{oee}+\hat\delta_{ooo}&=&I_1\,,\\
\hat\delta_{eee}-\hat\delta_{eoo}-\hat\delta_{oeo}+\hat\delta_{ooe}&=&R_2\,,\\
-\hat\delta_{eeo}+\hat\delta_{eoe}+\hat\delta_{oee}-\hat\delta_{ooo}&=&I_2\,,\\
\hat\delta_{eee}-\hat\delta_{eoo}+\hat\delta_{oeo}-\hat\delta_{ooe}&=&R_3\,,\\
\hat\delta_{eeo}-\hat\delta_{eoe}+\hat\delta_{oee}-\hat\delta_{ooo}&=&I_3\,,\\
\hat\delta_{eee}+\hat\delta_{eoo}-\hat\delta_{oeo}-\hat\delta_{ooe}&=&R_4\,,\\
-\hat\delta_{eeo}-\hat\delta_{eoe}+\hat\delta_{oee}+\hat\delta_{ooo}&=&I_4\,,
\end{eqnarray} 

\noindent where 

\begin{eqnarray}
\label{rone}
R_1&\equiv&{\rm Re}\,\hat\delta_{lmn}\,,\\
I_1&\equiv&{\rm Im}\,\hat\delta_{lmn}\,,\\
R_2&\equiv&{\rm Re}\,\hat\delta_{lm,N-n}\,,\\
I_2&\equiv&{\rm Im}\,\hat\delta_{lm,N-n}\,,\\
R_3&\equiv&{\rm Re}\,\hat\delta_{l,N-m,n}\,,\\
I_3&\equiv&{\rm Im}\,\hat\delta_{l,N-m,n}\,,\\
R_4&\equiv&{\rm Re}\,\hat\delta_{l,N-m,N-n}\,,\\
\label{ifour}
I_4&\equiv&{\rm Im}\,\hat\delta_{l,N-m,N-n}\,.
\end{eqnarray} 

\noindent Altogether, we get two separate systems of 4 equations and 4
unknowns. Written in matrix form, these systems are:
\begin{equation}
{\bf M}\left[\matrix{
\hat\delta_{eee}\cr\hat\delta_{eoo}\cr\hat\delta_{oeo}\cr\hat\delta_{ooe}\cr}
\right]=\left[\matrix{
R_1\cr R_2\cr R_3\cr R_4\cr}\right]\,,\qquad
{\bf M}\left[\matrix{
\hat\delta_{eeo}\cr\hat\delta_{eoe}\cr\hat\delta_{oee}\cr\hat\delta_{ooo}\cr}
\right]=\left[\matrix{
\phantom{-}I_1\cr -I_2\cr \phantom{-}I_3\cr -I_4\cr}\right]\,,
\end{equation}

\noindent where the matrix {\bf M} is given by
\begin{equation}
{\bf M}=\left[\matrix{
1 & \phantom{-}1 & \phantom{-}1 & \phantom{-}1 \cr
1 &           -1 &           -1 & \phantom{-}1 \cr
1 &           -1 & \phantom{-}1 &           -1 \cr
1 & \phantom{-}1 &           -1 &           -1 \cr}\right]\,.
\end{equation}

The inverse of the matrix ${\bf M}$ is simply ${\bf M}^{-1}={\bf M}/4$.
Hence, the $\hat\delta$'s are given by 
\begin{equation}
\label{sol1}
\left[\matrix{
\hat\delta_{eee}\cr\hat\delta_{eoo}\cr\hat\delta_{oeo}\cr\hat\delta_{ooe}\cr}
\right]={{\bf M}\over4}\left[\matrix{
R_1\cr R_2\cr R_3\cr R_4\cr}\right]\,,\qquad
\left[\matrix{
\hat\delta_{eeo}\cr\hat\delta_{eoe}\cr\hat\delta_{oee}\cr\hat\delta_{ooo}\cr}
\right]={{\bf M}\over4}\left[\matrix{
\phantom{-}I_1\cr -I_2\cr \phantom{-}I_3\cr -I_4\cr}\right]\,,
\end{equation}

\subsection{On a Face}

Consider now the case where one of the indicies, say $l$, is equal to
either 0 or $N/2$. These cases correspond to values of ${\bf k}$ located
on a face of the first octant in $k$-space. 
The problem becomes simpler. With $l=0$,
equation (\ref{mode1}) 
reduces to
\begin{eqnarray}
\hat\delta_{0mn}&=&\sum_{\alpha,\beta,\gamma=0}^{N-1}
\delta_{\alpha\beta\gamma}e^{2\pi im\beta/N}e^{2\pi in\gamma/N}\nonumber\\
\label{mode3}
&=&\sum_{\alpha,\beta,\gamma=0}^{N-1}\delta_{\alpha\beta\gamma}
\left(\cos{2\pi m\beta\over N}+i\sin{2\pi m\beta\over N}\right)
\left(\cos{2\pi n\gamma\over N}+i\sin{2\pi n\gamma\over N}\right)\,.
\end{eqnarray}

\noindent After expansion, this expression becomes
\begin{equation}
\label{d0mn}
\hat\delta_{0mn}=(\hat\delta_{ee}+\hat\delta_{oo})
+i(\hat\delta_{eo}+\hat\delta_{oe})\,,
\end{equation}

\noindent where
\begin{eqnarray}
\label{deltaee}
\hat\delta_{ee}&=&\sum_{\alpha,\beta,\gamma=0}^{N-1}\delta_{\alpha\beta\gamma}
\cos{2\pi m\beta\over N}\cos{2\pi n\gamma\over N}\,,\\
\hat\delta_{eo}&=&\sum_{\alpha,\beta,\gamma=0}^{N-1}\delta_{\alpha\beta\gamma}
\cos{2\pi m\beta\over N}\sin{2\pi n\gamma\over N}\,,\\
\hat\delta_{oe}&=&\sum_{\alpha,\beta,\gamma=0}^{N-1}\delta_{\alpha\beta\gamma}
\sin{2\pi m\beta\over N}\cos{2\pi n\gamma\over N}\,,\\
\label{deltaoo}
\hat\delta_{oo}&=&-\sum_{\alpha,\beta,\gamma=0}^{N-1}\delta_{\alpha\beta\gamma}
\sin{2\pi m\beta\over N}\sin{2\pi n\gamma\over N}\,.
\end{eqnarray}

Consider now the mode $\hat\delta_{0,N-m,n}$. We replace $m$ by $N-m$
in equation~(\ref{mode3}), and get
\begin{equation}
\label{mode3b}
\hat\delta_{0,N-m,n}=\sum_{\alpha,\beta,\gamma=0}^{N-1}
\delta_{\alpha\beta\gamma}e^{2\pi i(N-m)\beta/N}e^{2\pi in\gamma/N}
=\sum_{\alpha,\beta,\gamma=0}^{N-1}
\delta_{\alpha\beta\gamma}e^{2\pi i\beta}
e^{-2\pi im\beta/N}e^{2\pi in\gamma/N}\,.
\end{equation}

\noindent Since $\beta$ in an integer, the first exponential is always
unity, and equation~(\ref{mode3b}) reduces to
\begin{equation}
\label{mode3c}
\hat\delta_{0,N-m,n}
=\sum_{\alpha,\beta,\gamma=0}^{N-1}\delta_{\alpha\beta\gamma}
\left(\cos{2\pi m\beta\over N}-i\sin{2\pi m\beta\over N}\right)
\left(\cos{2\pi n\gamma\over N}+i\sin{2\pi n\gamma\over N}\right)\,.
\end{equation}

Comparing equations~(\ref{mode3}) and~(\ref{mode3c}), 
the only difference is a change of sign
of the first sine function. In equations~(\ref{deltaee})--(\ref{deltaoo}), 
that sine appears only
in the $\hat\delta$'s for which the first subscript is ``$o$.'' Hence, these
$\hat\delta$'s will change sign, and equation~(\ref{d0mn}) will becomes
\begin{equation}
\hat\delta_{0,N-m,n}=(\hat\delta_{ee}-\hat\delta_{oo})
+i(\hat\delta_{eo}-\hat\delta_{oe})\,,
\end{equation}

\noindent Similarly, we can easily show that replacing $n$ by $N-n$ results
in a change of sign of the $\hat\delta$'s for which the second subscript
is ``$o$.'' Hence, we get
\begin{eqnarray}
\label{d0mnn}
\hat\delta_{0m,N-n}&=&(\hat\delta_{ee}-\hat\delta_{oo})
+i(-\hat\delta_{eo}+\hat\delta_{oe})\,,\\
\hat\delta_{0,N-m,N-n}&=&(\hat\delta_{ee}+\hat\delta_{oo})
+i(-\hat\delta_{eo}-\hat\delta_{oe})\,.
\end{eqnarray}

\begin{figure}[t]
\vskip-0.5in
\hskip0.3in
\includegraphics[width=6.2in]{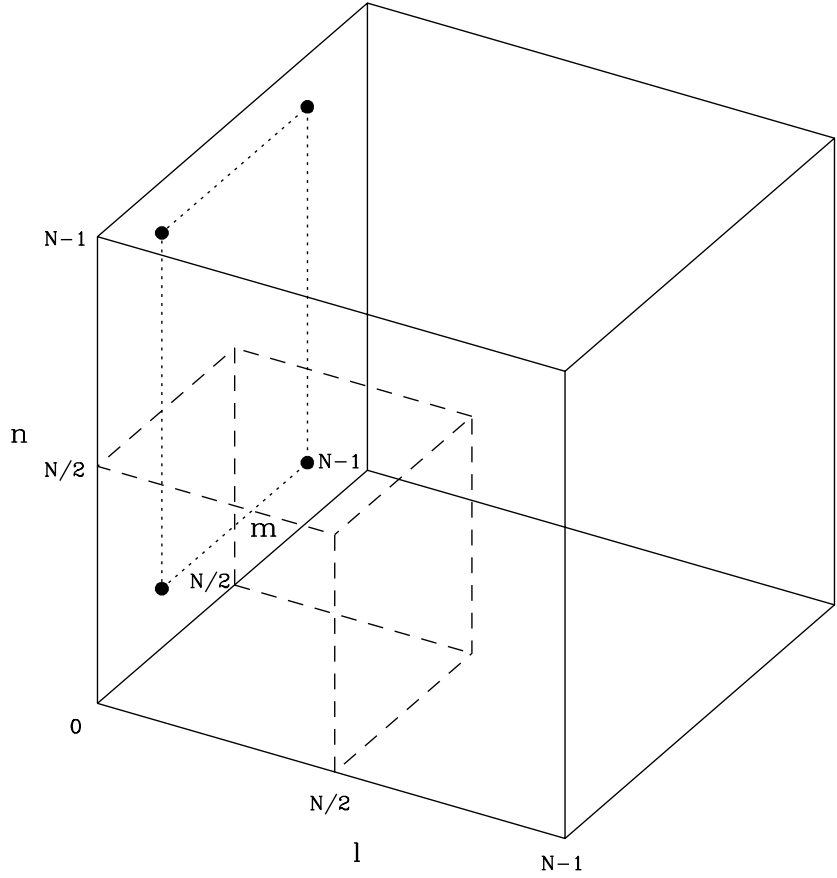}
\vskip-0.7in
\caption{Representation of the system in $k$-space. The dots
indicate the particular mode $(l,m,n)=(0,15,8)$ and the three related
modes.}
\label{cube2}
\end{figure}

\noindent These 4 modes are located at the verticies of a rectangle
in $k$-space, as shown in Figure~\ref{cube2}.
They form two pairs of complex conjugates,
\begin{eqnarray}
\hat\delta_{0mn}&=&\hat\delta_{0,N-m,N-n}^*\,,\\
\hat\delta_{0m,N-n}&=&\hat\delta_{0,N-m,n}^*\,.
\end{eqnarray}

\noindent By equating these expressions with 
equations~(\ref{d0mn}) and (\ref{d0mnn}), and
considering the real and imaginary parts separately, we get 4 equations,
\begin{eqnarray}
\hat\delta_{ee}+\hat\delta_{oo}&=&R_1\,,\\
\hat\delta_{eo}+\hat\delta_{oe}&=&I_1\,,\\
\hat\delta_{ee}-\hat\delta_{oo}&=&R_2\,,\\
-\hat\delta_{eo}+\hat\delta_{oe}&=&I_2\,,
\end{eqnarray}

\noindent where
\begin{eqnarray}
\label{roneb}
R_1&\equiv&{\rm Re}\,\hat\delta_{0mn}\,,\\
I_1&\equiv&{\rm Im}\,\hat\delta_{0mn}\,,\\
R_2&\equiv&{\rm Re}\,\hat\delta_{0m,N-n}\,,\\
I_2&\equiv&{\rm Im}\,\hat\delta_{0m,N-n}\,.
\label{itwo}
\end{eqnarray}

\noindent Again, the numbers $R_1$, $I_1$, $R_2$, and $I_2$ are
determined from equations~(\ref{real}) and~(\ref{imaginary}).
The solutions are
\begin{eqnarray}
\label{deltaee2}
\hat\delta_{ee}&=&{R_1+R_2\over2}\,,\\
\hat\delta_{oo}&=&{R_1-R_2\over2}\,,\\
\hat\delta_{eo}&=&{I_1-I_2\over2}\,,\\
\label{deltaoo2}
\hat\delta_{oe}&=&{I_1+I_2\over2}\,.
\end{eqnarray}

Consider now the case $l=N/2$. Equation (\ref{mode1}) reduces to
\begin{eqnarray}
\hat\delta_{N/2,mn}&=&\sum_{\alpha,\beta,\gamma=0}^{N-1}
\delta_{\alpha\beta\gamma}e^{\pi il\alpha}
e^{2\pi im\beta/N}e^{2\pi in\gamma/N}\nonumber\\
&=&\sum_{\alpha,\beta,\gamma=0}^{N-1}\delta_{\alpha\beta\gamma}
(-1)^\alpha
\left(\cos{2\pi m\beta\over N}+i\sin{2\pi m\beta\over N}\right)
\left(\cos{2\pi n\gamma\over N}+i\sin{2\pi n\gamma\over N}\right)\,.
\end{eqnarray}

\noindent After expansion, this expression becomes
\begin{equation}
\hat\delta_{0mn}=(\hat\delta_{ee}+\hat\delta_{oo})
+i(\hat\delta_{eo}+\hat\delta_{oe})\,,
\end{equation}

\noindent where
\begin{eqnarray}
\hat\delta_{ee}&=&\sum_{\alpha,\beta,\gamma=0}^{N-1}\delta_{\alpha\beta\gamma}
(-1)^\alpha
\cos{2\pi m\beta\over N}\cos{2\pi n\gamma\over N}\,,\\
\hat\delta_{eo}&=&\sum_{\alpha,\beta,\gamma=0}^{N-1}\delta_{\alpha\beta\gamma}
(-1)^\alpha
\cos{2\pi m\beta\over N}\sin{2\pi n\gamma\over N}\,,\\
\hat\delta_{oe}&=&\sum_{\alpha,\beta,\gamma=0}^{N-1}\delta_{\alpha\beta\gamma}
(-1)^\alpha
\sin{2\pi m\beta\over N}\cos{2\pi n\gamma\over N}\,,\\
\hat\delta_{oo}&=&-\sum_{\alpha,\beta,\gamma=0}^{N-1}\delta_{\alpha\beta\gamma}
(-1)^\alpha
\sin{2\pi m\beta\over N}\sin{2\pi n\gamma\over N}\,.
\end{eqnarray}

\begin{figure}[t]
\vskip-0.5in
\hskip0.3in
\includegraphics[width=6.2in]{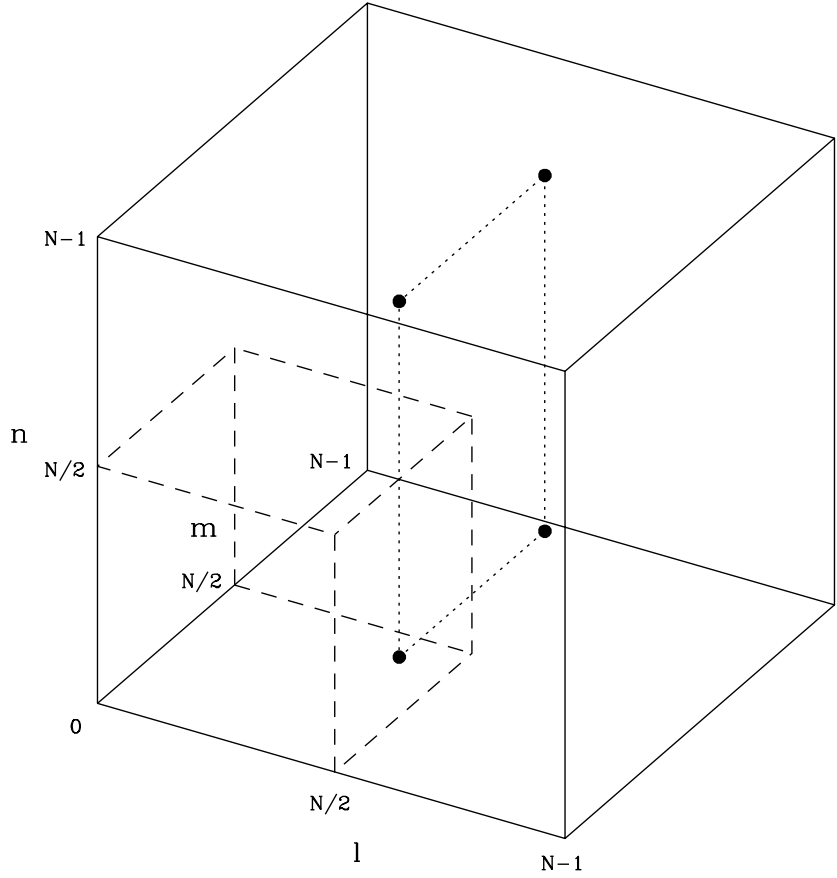}
\vskip-0.7in
\caption{Representation of the system in $k$-space. The dots
indicate the particular mode $(l,m,n)=(32,15,8)$, where $32=N/2$,
and the three related modes.}
\label{cube2b}
\end{figure}

We find the same expressions as for the case $l=0$, the only differences
being the extra factor of $(-1)^\alpha$ in the definitions of the
$\hat\delta$'s. Hence, 
the solutions~(\ref{deltaee2})--(\ref{deltaoo2}) are still valid in this case.
These modes are shown in Figure~\ref{cube2b}.

We have only considered the cases when the first index, $l$, is either
0 or $N/2$, but these results can be generalized to the other indicies
$m$ and $n$ as well, since the entire problem has cubic symmetry.

\subsection{On an Edge}

Consider now the case when two of the indicies, say $l$ and $m$, are
equal to either 0 or $N/2$. These cases corresponds to values of
${\bf k}$  located on an edge of the first octant in $k$-space. With $l=m=0$,
equation (\ref{mode1}) reduces to
\begin{equation}
\label{mode4}
\hat\delta_{00n}=\sum_{\alpha,\beta,\gamma=0}^{N-1}
\delta_{\alpha\beta\gamma}e^{2\pi in\gamma/N}
=\sum_{\alpha,\beta,\gamma=0}^{N-1}\delta_{\alpha\beta\gamma}
\left(\cos{2\pi in\gamma\over N}+i\sin{2\pi in\gamma\over N}\right)\,.
\end{equation}

\noindent This expression becomes
\begin{equation}
\label{d00n}
\hat\delta_{00n}=\hat\delta_e+i\hat\delta_o\,,
\end{equation}

\noindent where
\begin{eqnarray}
\label{deltae}
\hat\delta_e&=&\sum_{\alpha,\beta,\gamma=0}^{N-1}\delta_{\alpha\beta\gamma}
\cos{2\pi n\gamma\over N}\,,\\
\label{deltao}
\hat\delta_o&=&\sum_{\alpha,\beta,\gamma=0}^{N-1}\delta_{\alpha\beta\gamma}
\sin{2\pi n\gamma\over N}\,.
\end{eqnarray}

Consider now the mode $\hat\delta_{00,N-n}$. We replace $n$ by $N-n$
in equation~(\ref{mode4}), and get
\begin{equation}
\label{mode4b}
\hat\delta_{00,N-n}=\sum_{\alpha,\beta,\gamma=0}^{N-1}
\delta_{\alpha\beta\gamma}e^{2\pi i(N-n)\gamma/N}
=\sum_{\alpha,\beta,\gamma=0}^{N-1}
\delta_{\alpha\beta\gamma}e^{2\pi i\gamma}
e^{-2\pi in\gamma/N}\,.
\end{equation}

\noindent Since $\gamma$ in an integer, the first exponential is always
unity, and equation~(\ref{mode4b}) reduces to
\begin{equation}
\label{mode4c}
\hat\delta_{0,0,N-n}
=\sum_{\alpha,\beta,\gamma=0}^{N-1}\delta_{\alpha\beta\gamma}
\left(\cos{2\pi n\gamma\over N}-i\sin{2\pi n\gamma\over N}\right)\,.
\end{equation}

Comparing equations~(\ref{mode4}) 
and~(\ref{mode4c}), the only difference is a change of sign
of the sine function. In equations~(\ref{deltae}) 
and~(\ref{deltao}), that sine appears only
in the expression for $\hat\delta_o$. Hence, equation~(\ref{d00n}) becomes
\begin{equation}
\label{d00nn}
\hat\delta_{00,N-n}=\hat\delta_e-i\hat\delta_o\,,
\end{equation}

\begin{figure}[t]
\vskip-0.5in
\hskip0.3in
\includegraphics[width=6.2in]{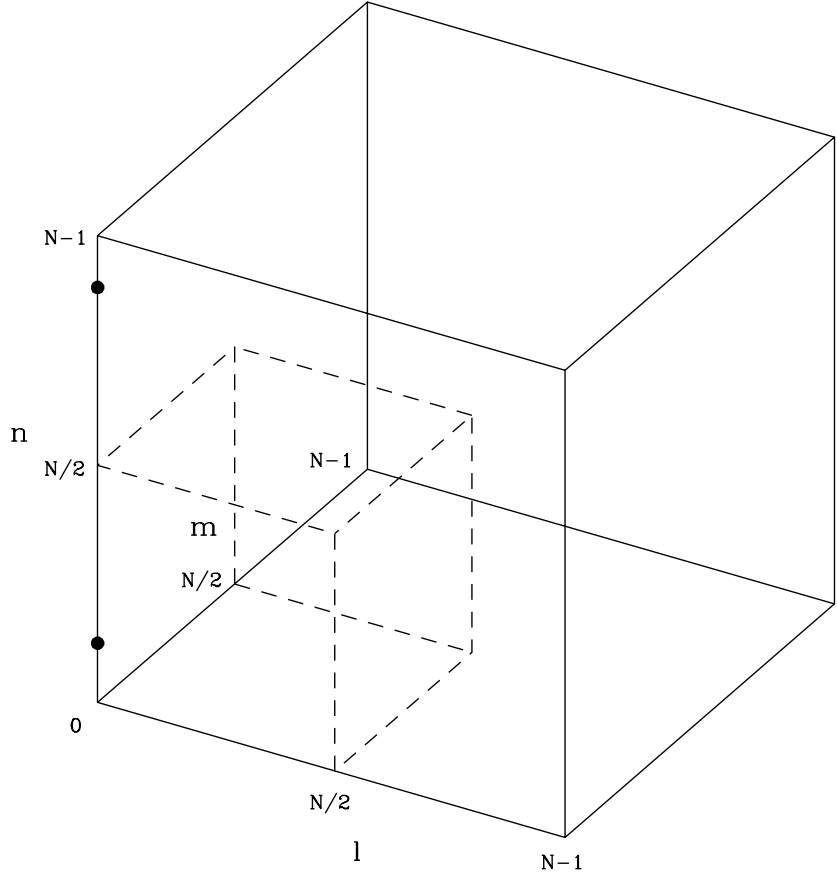}
\vskip-0.7in
\caption{Representation of the system in $k$-space. The dots
indicate the particular mode $(l,m,n)=(0,0,8)$ and its related
mode.}
\label{cube3}
\end{figure}

\noindent These two modes are complex conjugates,
\begin{eqnarray}
\label{d00n2}
\hat\delta_{00n}&=&\hat\delta_{00,N-n}^*\,.
\end{eqnarray}

\noindent They are shown in Figure~\ref{cube3}.
By equating equation~(\ref{d00n}) and (\ref{d00n2}), and
considering the real and imaginary parts separately, we get
\begin{eqnarray}
\label{deltae2}
\hat\delta_e&=&R_1\,,\\
\label{deltao2}
\hat\delta_o&=&I_1\,,
\end{eqnarray}

\noindent where
\begin{eqnarray}
R_1&\equiv&{\rm Re}\,\hat\delta_{00n}\,,\\
\label{ione}
I_1&\equiv&{\rm Im}\,\hat\delta_{00n}\,.
\end{eqnarray}

\noindent The numbers $R_1$ and $I_1$  are
determined from equations~(\ref{real}) and~(\ref{imaginary}).

\begin{figure}[t]
\vskip-0.5in
\hskip0.3in
\includegraphics[width=6.2in]{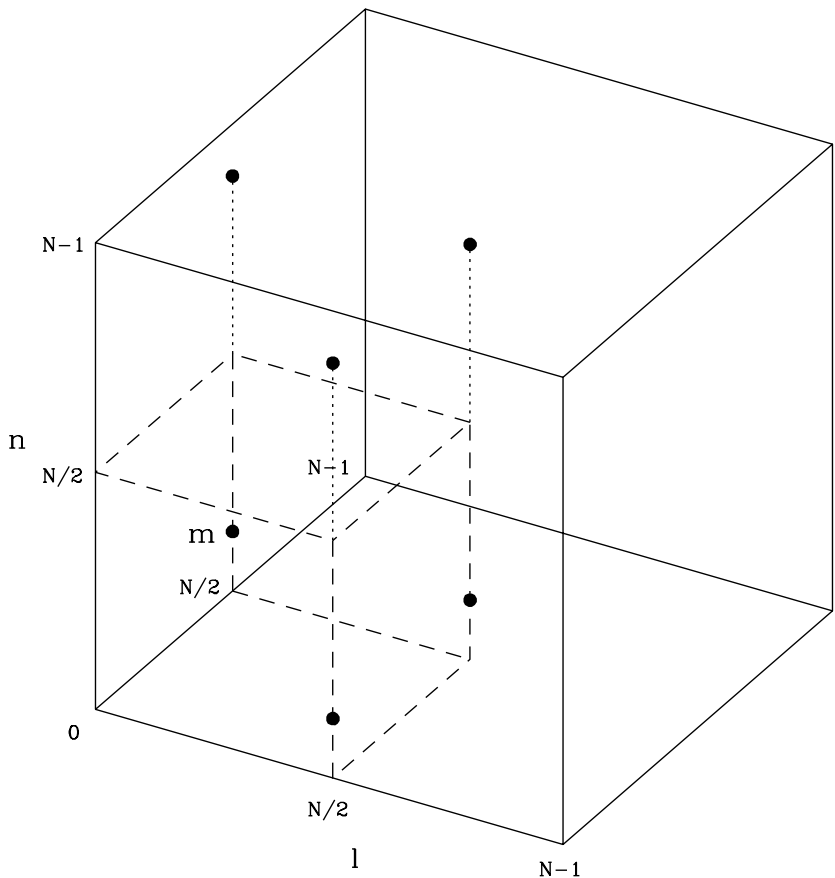}
\vskip-0.7in
\caption{Representation of the system in $k$-space. The dots
indicate the particular modes $(l,m,n)=(0,32,8)$, $(32,0,8)$, and
$(32,32,8)$, where $32=N/2$, and their related modes.}
\label{cube3b}
\end{figure}

Consider now the case $l=N/2$, $m=0$, the case $l=0$, $m=N/2$,
and the case $l=m=N/2$. We can easily show that the
solutions~(\ref{deltae2}) and~(\ref{deltao2}) 
are still valid, using the same approach as in \S3.2.
Again, the only difference will be extra factors
of $(-1)^\alpha$ in the definitions of $\hat\delta_e$ and $\hat\delta_o$.
These modes are shown in Figure~\ref{cube3b}.
Using the cubic symmetry, we can then show that the solutions~(\ref{deltae2}) 
and~(\ref{deltao2})
applies to all cases for which two of the three indicies $l$, $m$, $n$
are equal to 0 or $N/2$ (12 combinations).

\subsection{In a Corner}

\begin{figure}[t]
\vskip-0.5in
\hskip0.3in
\includegraphics[width=6.2in]{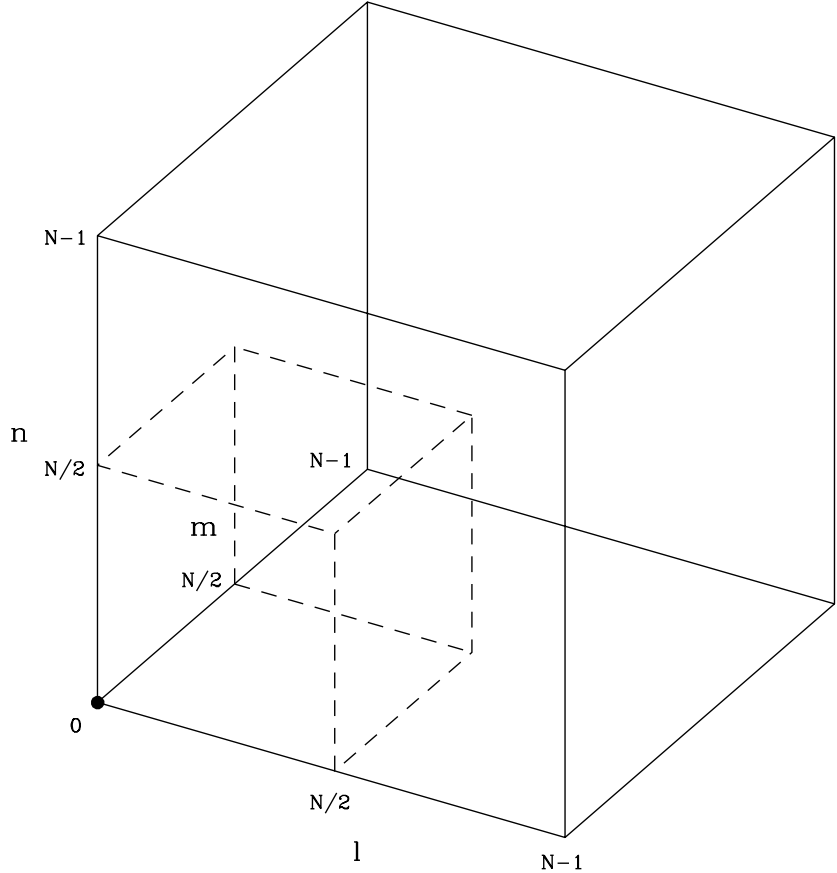}
\vskip-0.7in
\caption{Representation of the system in $k$-space. The dot
indicates the particular mode $(l,m,n)=(0,0,0)$. For that particular
mode, we ignore equation~(\ref{deltau}) and set $\delta_u=0$ instead.}
\label{cube4}
\end{figure}

Finally, we consider the cases when all indicies are equal to 0 or
$N/2$. These cases corresponds to values of
${\bf k}$  located in a corner of the first octant in $k$-space.
With $l=m=n=0$, equation (\ref{mode1}) reduces to
\begin{equation}
\label{deltau}
\hat\delta_{000}=\sum_{\alpha,\beta,\gamma=0}^{N-1}
\delta_{\alpha\beta\gamma}\equiv\delta_u\,,
\end{equation}

\noindent where the subscript $u$ stands for ``uniform,'' since the
mode 000 corresponds to a null wavenumber, or an infinite wavelength.
Notice that since $\delta_{\alpha\beta\gamma}$ is real, $\delta_u$ is
real as well. Hence, for this mode, there is no imaginary part, and
$\delta_u$ is determined form equation~(\ref{real}).
This mode is shown in Figure~\ref{cube4}.

Consider now the case $l=N/2$, $m=0$, $n=0$. 
Again, we can easily show that the
solution~(\ref{deltau}) is still valid, using the same approach as in 
\S\S3.2 and~3.3. The only difference will be an extra factor
of $(-1)^\alpha$ in the expression for $\delta_u$.
Using the cubic symmetry, we can then show that the solution~(\ref{deltau})
applies to all cases for which the three indicies $l$, $m$, $n$
are equal to 0 or $N/2$ (8 combinations).

\begin{figure}[t]
\vskip-0.5in
\hskip0.3in
\includegraphics[width=6.2in]{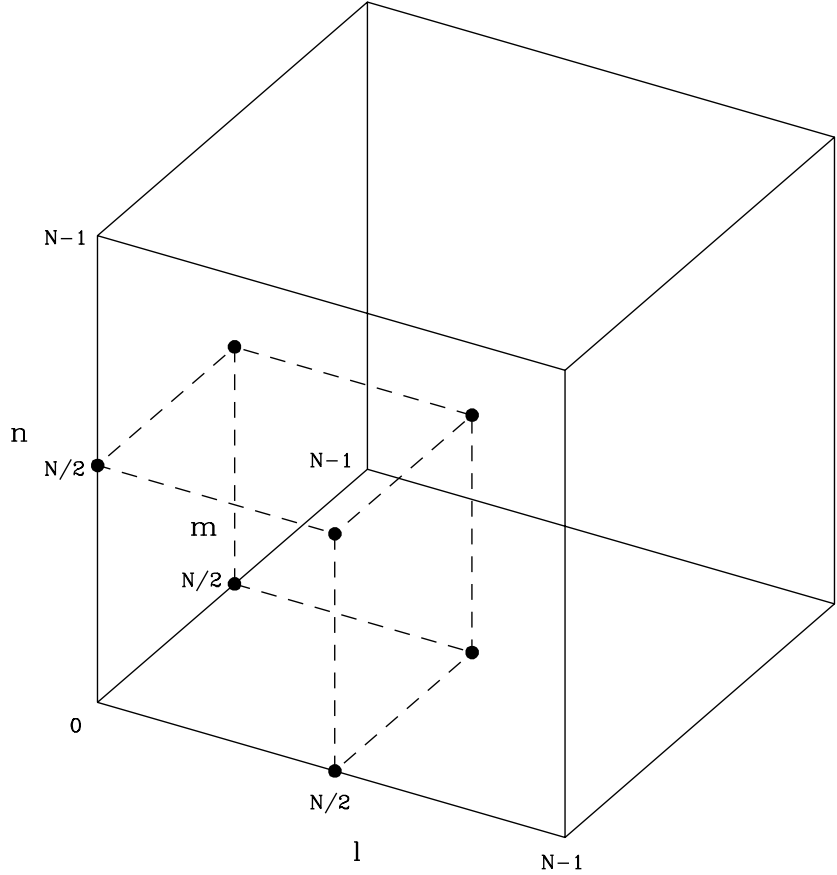}
\vskip-0.7in
\caption{Representation of the system in $k$-space. The dots
indicate the particular modes $(l,m,n)$ located at corners of the
first octant, excluding the mode $(0,0,0)$.
For these modes, equation~(\ref{deltau}) applies.}
\label{cube4b}
\end{figure}

Notice that there is a fundamental difference between the mode
$\hat\delta_{000}$ and the other 7 modes,
$\hat\delta_{00,N/2}$, $\hat\delta_{0,N/2,0}$, $\ldots$,
$\hat\delta_{N/2,N/2,N/2}$, shown in Figures~\ref{cube4} and
\ref{cube4b}, respectively. The mode $\hat\delta_{000}$ represents
a perturbation of infinite wavelength, that is, a constant. Clearly
that constant must be zero, otherwise the mean value of $\delta({\bf r})$
integrated over the entire volume would be nonzero, and this would
violate the assumption that the volume contains a fair sample of the
universe. Therefore, for the mode $\hat\delta_{000}$,
{\it and that mode only},
we do not compute the amplitude $|\hat\delta_{000}|$ from the
power spectrum, but instead set that amplitude equal to zero.

\subsection{Putting it All Together}

We can now count the number of independent quantities necessary to represent
all the density harmonics. Consider first the modes $\hat\delta_{lmn}$ for
which the indicies are neither 0 nor $N/2$. Excluding these values, each
index can take $N-2$ different values, which gives us $(N-2)^3$ modes.
As we showed in \S3.1, these modes come in groups of 8, and 
within each
group the complex values of the harmonics can be expressed as combinations
of 8 real numbers $\hat\delta_{eee}$, $\hat\delta_{eeo}$, $\ldots$,
$\hat\delta_{ooo}$. Hence, it takes a total of $(N-2)^3$ real numbers to
represent these modes.

Consider now the modes for which one of the indicies is equal to 0 or $N/2$.
There are 6 possibilities, and for each one, the remaining two indicies can
take $N-2$ values each (all values except 0 and $N/2$). This gives us 
$6(N-2)^2$ modes. As we showed in \S3.2, these modes come in 
groups of four, and within each
group the complex values of the harmonics can be expressed as combinations
of four real numbers $\hat\delta_{ee}$, $\hat\delta_{eo}$,
$\hat\delta_{oe}$, and $\hat\delta_{oo}$. 
Hence, it takes a total of $6(N-2)^2$ real numbers to
represent these modes.

Next, consider the modes for which two of the indicies are equal to 0 or $N/2$.
There are 12 possibilities, and for each one, the remaining index can
take $N-2$ values (all values except 0 and $N/2$). This gives us 
$12(N-2)$ modes. As we showed in \S3.3, the amplitude these modes form
complex conjugates pairs, and each pair can be represented
by two real numbers $\hat\delta_e$, and $\hat\delta_o$.
Hence, it takes a total of $12(N-2)$ real numbers to
represent these modes.

Finally, 
consider the modes for which all three indicies are equal to 0 or $N/2$.
There are 8 such modes. As we showed in \S3.4, these modes are
real, hence it takes 8 real numbers to represent them. The total number
of variables necessary to represent all the density harmonics is
therefore
\begin{equation}
(N-2)^3+6(N-2)^2+12(n-2)+8=N^3\,.
\end{equation}

\noindent It takes $N^3$ numbers to represent the Fourier
transform of a cube $N\times N\times N$ of real numbers, in spite of the fact
that the Fourier transform is complex. Indeed, it is common
for Fast Fourier Transform (FFT) subroutines to take a tridimensional
array of number and to overwrite that array with the Fourier transform.
This is the case for the FFT subroutines of
{\sl Numerical Recipes} \citep{numrec}. 
When these subroutines compute the Fourier
transform of a cube $N\times N\times N$, the results are written
in the same cube, and {\it the actual numbers stored in the cube are
precisely the $\hat\delta_{xxx}$'s,  $\hat\delta_{xx}$'s,  $\hat\delta_x$'s,
and $\delta_u$'s derived in this section}. Hence it is possible to generate
the density harmonics directly in $k$-space, using the above formulae, store
these numbers at the appropriate locations inside an $N\times N\times N$
cubic array, and then invoke the {\sl Numerical Recipes} inverse FFT
subroutines to generate the density field. 

The {\sl Numerical Recipes}
convention for storing the density harmonics is the following: Consider
a 3D array {\tt A}, with indicies running from 0 to $N-1$. We loop over
all modes located in the first octant in $k$-space: $0\leq l,m,n\leq N/2$.

\begin{enumerate}

\item For modes with $0<l,m,n<N/2$ (the general case), the numbers
$\delta_{eee}$, $\delta_{eeo}$, $\ldots$, $\delta_{ooo}$ are stored
in a $2\times2\times2$ cube located at {\tt A(2l,2m,2n)}, {\tt A(2l,2m,2n+1)},
$\ldots$, {\tt A(2l+1,2m+1,2n+1)}.

\item For modes with $l=0$, the numbers $\delta_{ee}$, $\delta_{eo}$,
$\delta_{oe}$, $\delta_{oo}$ are stored at 
{\tt A(0,2m,2n)}, {\tt A(0,2m,2n+1)}, {\tt A(0,2m+1,2n)}, 
and {\tt A(0,2m+1,2n+1)}, respectively. For modes with $l=N/2$, they are stored
at {\tt A(1,2m,2n)}, {\tt A(1,2m,2n+1)}, {\tt A(1,2m+1,2n)}, 
and {\tt A(1,2m+1,2n+1)}, respectively. This is easily generalized to the
other faces ($m=0$, $m=N/2$, $n=0$, $n=N/2$).

\item For modes with $l=m=0$, the numbers $\delta_e$ and $\delta_o$
are stored at {\tt A(0,0,2n)} and {\tt A(0,0,2n+1)}, respectively.
For modes with $l=0$, $m=N/2$, they are stored
are stored at {\tt A(0,1,2n)} and {\tt A(0,1,2n+1)}, respectively.
This is easily generalized to the other edges.

\item The number $\delta_u$ is stored in a $2\times2\times2$ cube located
in the corner of the array, at {\tt A(0,0,0)}, {\tt A(0,0,1)}, $\ldots$,
{\tt A(1,1,1)}. Then, the value of {\tt A(0,0,0)} is set to 0, since this
represents the mode $(0,0,0)$.

\end{enumerate}

\section{CALCULATION OF THE DENSITY FIELD}

\subsection{Eulerian Representation}

With the expressions derived in \S2, we have all the
ingredients necessary to compute the density field on a grid. The steps are 
the following:

\begin{enumerate}

\item Choose a particular power spectrum $P(k)$, a box size $L_{\rm box}$,
and a grid size $N$. This determines the fundamental wavenumber 
$k_0=2\pi/L_{\rm box}$. The allowed modes are given by 
equation~(\ref{kallowed}).

\item For all modes with $l,m,n\neq0,N/2$, group these modes in groups
of 8, and for each group, calculate the quantities 
$\delta_{eee}$, $\delta_{eeo}$, $\ldots$, $\delta_{ooo}$ using 
equations~(\ref{rone})--(\ref{ifour}) and~(\ref{sol1}).
The quantities $R_1$, $I_1$, $\ldots$, $I_4$ are determined from 
equations~(\ref{real}) and~(\ref{imaginary}).

\item For all modes located on a face (one of the indicies $l,m,n$ equal
to 0 or $N/2$), group these modes in groups
of four, and for each group, calculate the quantities 
$\delta_{ee}$, $\delta_{eo}$, $\delta_{oe}$, $\delta_{oo}$ using 
equations~(\ref{roneb})--(\ref{itwo}) and~(\ref{deltaee2})--(\ref{deltaoo2}).

\item For all modes located on an edge (two of the indicies $l,m,n$ equal
to either  0 or $N/2$), group these modes in groups
of two, and for each group, calculate the quantities 
$\delta_e$ and $\delta_o$, using 
equations~(\ref{deltae2})--(\ref{itwo}).

\item For the modes located in a corner (all indicies $l,m,n$ equal
to either  0 or $N/2$), calculate the quantity
$\delta_u$ using equation~(\ref{deltau}). For the mode
$(l,m,n)=(0,0,0)$, replace the value of $\delta_u$ by 0.

\item Store the quantities $\delta_{eee}$, $\ldots$, $\delta_{ooo}$,
$\delta_{ee}$, $\delta_{eo}$, $\delta_{oe}$, $\delta_{oo}$, $\delta_e$,
$\delta_o$, $\delta_u$ in a 3D, $N\times N\times N$ array, at the proper 
locations. These depends on the convention used by the FFT subroutines.

\item Compute the inverse FFT of the 3D array. The result will be the
density field $\delta(\rbf)$.

\end{enumerate}

\subsection{Lagrangian Representation}

In the Lagrangian representation, the density field is represented by a
distribution of equal-mass particles. We start by laying down 
$N_p\times N_p\times N_p$ particles on a cubic grid with grid spacing 
$d=L_{\rm box}/N_p$
inside the computational volume $V_{\rm box}$. The mass of the particles
are given by $M=\bar\rho V_{\rm box}/N_p^3$, such that the mean density
inside the box is equal to the mean background density $\bar\rho$. 
The particles are then displaced in order to represent the initial density 
field. This approach is valid only in the {\it linear regime}, defined by
$|\delta(\rbf)|\ll1$. In this regime, the particle displacements 
$\Delta\rbf$ are significantly smallar than the initial separation $d$
between the particles, and can be computed using the 
Zel'dovich approximation \citep{zeldovich70}. The
displacements are given by
\begin{equation}
\label{direct}
\Delta\rbf_j=-{i\over N^3}\sum_\kbf{\hat\delta_\kbf\kbf\over k^2}
e^{-i\kbf\cdot\rbf_j}\,,
\end{equation}

\noindent where $\rbf_j$ is the position of particle $j$ before it is
displaced, and $\rbf_j+\Delta\rbf_j$ is the position after. Notice
that the reality condition~(\ref{reality}) ensures that $\Delta\rbf_j$
is real.

This method is straightforward, but can rapidly become unpractical.
Equation~(\ref{direct}) involves a sum over $N^3$ modes, and that sum
must be performed for each of the $N_p^3$ particles. Since in typical
cosmological simulations $N_p$ is chosen to be $N/2$, the number of 
operations scales like $N^6$. If it takes 5 minutes
to set up initial conditions with $64^3$ particles, it will take
5.3 hours for $128^3$ particles, 2 weeks for $256^3$ particles,
and 2.5 years for $512^3$ particles! An alternative method,
which scales like $N^3$, was proposed by \citet{edfw85}. 
Essentially, this approach uses the fact that 
the displacement of each particle is
proportional to its peculiar acceleration, that can be
calculated with a N-body simulation algorithm such as PM
(Particle-Mesh) or $\rm P^3M$ (Particle-Particle/Particle-Mesh). 
With this method, the number of operations scales roughly as $N^3$.
I refer the reader to \citet{he81} and \citet{edfw85} for details.
It is worth noting that, unlike equation~(\ref{direct}), the approach
of \citet{edfw85} is approximative, and the high-frequency modes, near
the Nyquist frequency $k=(N_p/2)k_0$, are often poorly represented
by the particle distribution.

\section{FILTERING}

Our next task in to filter the density field at some scale $s$. The choice of
scale must obey two conditions: $s\gg\Delta$ and $s\ll L_{\rm box}$.
The first condition is required by the discreteness of the grid
and the second by the assumption of periodic boundary conditions.
The filtered density field $\delta_s(\rbf)$ at scale $s$ is given by
\begin{equation}
\label{deltaf}
\delta_s(\rbf)=\int_{V_{\rm box}}\delta(\rbf')K_s(\rbf-\rbf')d^3r'\,,
\end{equation}

\noindent where $K_s$ is the filter function. We will use a Gaussian filter
given by
\begin{equation}
K_s(\xbf)={e^{-x^2/2s^2}\over(2\pi)^{3/2}s^3}\,.
\end{equation}

\noindent This filter function satisfied the normalization condition,
\begin{equation}
\int_{V_{\rm box}}K_s(\rbf)d^3r=1\,,
\end{equation}

\noindent as long as $s\ll L_{\rm box}$.

It is well known that filtering in real space is equivalent to a 
multiplication in $k$-space. However, it is useful to redo the derivation,
to ensure that we have all the correct factors of $2\pi$, $N^3$, and so on.
First, we express the filter as an inverse Fourier transform,
\begin{equation}
\label{fft3}
K_s(\rbf-\rbf')={1\over N^3}\sum_{\kbf'}
\hat K_s(\kbf')e^{-i\kbf'\cdot(\rbf-\rbf')}\,.
\end{equation}

\noindent We substitute equations~(\ref{fft2}) and~(\ref{fft3}) in
equation~(\ref{deltaf}), and get
\begin{eqnarray}
\label{deltaf2}
\delta_s(\rbf)&=&{1\over N^6}\int_{V_{\rm box}}
\sum_\kbf\hat\delta(\kbf)e^{-i\kbf\cdot\rbf}
\sum_{\kbf'}\hat K_s(\kbf')e^{-i\kbf'\cdot(\rbf-\rbf')}d^3r'\nonumber\\
&=&{1\over N^6}\sum_\kbf\sum_{\kbf'}\hat\delta(\kbf)\hat K_s(\kbf')
e^{-i\kbf\cdot\rbf}\int_{V_{\rm box}}e^{i(\kbf'-\kbf)\cdot\rbf'}d^3r'\,.\\
\end{eqnarray}

\noindent The integral is equal to $V_{\rm box}\delta_{\kbf,\kbf'}$ (see
eq.~[\ref{dkk}]), and we use the Kronecker $\delta$ to eliminate the sum
over $\kbf'$. We get
\begin{equation}
\label{deltaf3}
\delta_s(\rbf)={V_{\rm box}\over N^6}
\sum_\kbf\hat\delta(\kbf)\hat K_s(\kbf)e^{-i\kbf\cdot\rbf}\,.
\end{equation}

We now need an expression for $\hat K_s(\kbf)$. This function is the
Fourier transform of the filter,
\begin{equation}
\hat K_s(\kbf)=\sum_\xbf K_s(\xbf)e^{i\kbf\cdot\xbf}
={1\over(2\pi)^{3/2}s^3}\sum_\xbf e^{-x^2/2s^2}e^{i\kbf\cdot\xbf}\,.
\end{equation}

\noindent We rewrite this expression as
\begin{equation}
\hat K_s(\kbf)={1\over(2\pi)^{3/2}s^3}\left({N^3\over V_{\rm box}}\right)
\sum_\xbf e^{-x^2/2s^2}e^{i\kbf\cdot\xbf}\left({V_{\rm box}\over N^3}\right)\,.
\end{equation}

\noindent The factor $V_{\rm box}/N^3=\Delta^3$ 
represents the volume element around
each point $\xbf$ in the $N\times N\times N$ grid. 
Since we assume $s\gg\Delta$, we can approximate the sum as an integral
over the volume of the box (or, equivalently, regard the sum as a numerical
approximation for the integral). Hence,
\begin{equation}
\hat K_s(\kbf)={1\over(2\pi)^{3/2}s^3}\left({N^3\over V_{\rm box}}\right)
\int_{V_{\rm box}} e^{-x^2/2s^2}e^{i\kbf\cdot\xbf}d^3x\,.
\end{equation}

\noindent Since we assume periodic boundary conditions, we are free to locate
the origin anywhere inside the box. For instance, we can locate it in the 
center of the box. Since $s\ll L_{\rm box}$, the integrant becomes negligible
at the edge of the box. We can then extend the integration domain to
all space,
\begin{equation}
\label{intall}
\hat K_s(\kbf)={1\over(2\pi)^{3/2}s^3}\left({N^3\over V_{\rm box}}\right)
\int_{\hbox{all space}} e^{-x^2/2s^2}e^{i\kbf\cdot\xbf}d^3x\,,
\end{equation}

\noindent where this expression {\it no longer assumes boundary conditions\/}.
The integral in equation~(\ref{intall}) can be found in any textbook
of Fourier transforms,
\begin{equation}
\int_{\hbox{all space}} e^{-x^2/2s^2}e^{i\kbf\cdot\xbf}d^3x
=(2\pi)^{3/2}s^3e^{-(ks)^2/2}\,.
\end{equation}

\noindent Hence,
\begin{equation}
\hat K_s(\kbf)={N^3\over V_{\rm box}}e^{-(ks)^2/2}\,.
\end{equation}

\noindent We substitute this expression in equation~(\ref{deltaf3}), and get
\begin{equation}
\label{deltaf4}
\delta_s(\rbf)={1\over N^3}
\sum_\kbf\hat\delta(\kbf)e^{-(ks)^2/2}e^{-i\kbf\cdot\rbf}\,.
\end{equation}

\noindent Hence, to obtain a filtered density field, we generate the density
harmonics using the method described in \S3,
and then multiply them by the factor $e^{-k^2s^2/2}$ {\it before\/}
taking the inverse Fourier transform.

\section{SUMMARY}

This paper presents in great detail the techniques used for generating 
Gaussian density fields. These techniques are well-known among experts
in cosmological numerical simulations, but the specific details of the
implementation are often difficult to find in the literature. Also,
the notation tends to vary significantly from one author to another.
The consequences is that any new researcher moving into this field
has to either spend a great deal of effort rederiving all the
technical details, or else rely on existing codes and
use them as black boxes. The goal of this document is to improve the
situation by presenting in a comprehensive form the basic theory
behind the generation of Gaussian random fields.

\acknowledgments

I am very thankful to Yehuda Hoffman, Patrick McDonald,
Matthew Pieri, C\'edric Grenon, and Matthew Craig 
for reading this manuscript and
making valuable comments. This work was supported by the Canada Research
Chair program and NSERC.

\appendix

\section{FOURIER TRANSFORM OF THE TOP HAT}

Consider the following integral,
\begin{equation}
\label{a1}
I=\int_{\rm sph(0)}d^3y\,e^{-i\kbf\cdot\ybf}\,,
\end{equation}

\noindent where the domain of integration is a sphere of radius $R$ centered
at the origin. We consider a spherical coordinate system centered at the 
origin, with the $z$-axis pointing in the direction of $\kbf$. 
Equation~(\ref{a1}) becomes
\begin{equation}
I=\int_0^{2\pi}d\phi\int_0^\pi d\theta
\int_0^R dy\,(y^2\sin\theta)e^{-iky\cos\theta}\,,
\end{equation}

\noindent where $k\equiv|\kbf|$, $y\equiv|\ybf|$, and $\theta$ is the
angle between $\kbf$ and $\ybf$. The integrations over $\phi$ and $\theta$
are trivial. We get
\begin{equation}
I=2\pi\int_0^R dy\,y^2\left[{e^{-iky\cos\theta}\over iky}\right]_0^\pi
=2\pi\int_0^R dy\,y^2\left[{e^{iky}-e^{-iky}\over iky}\right]
=4\pi\int_0^R dy\,{y\sin ky\over k}\,.
\end{equation}

\noindent The integral over $y$ is now trivial. We get
\begin{equation}
I={4\pi\over k^3}(\sin kR-kR\cos kR)={4\pi R^3\over u^3}(\sin u-u\cos u)\,,
\end{equation}

\noindent where $u=kR$.

%


\clearpage

\begin{thebibliography}{}

\bibitem[Bunn \& White(1997)]{bw97} Bunn, E. F., \& White, M. 1997,
        \apj, 480, 6

\bibitem[Efstathiou et al.(1985)]{edfw85}
        Efstathiou, G., Davis, M., Frenk, C. S., \& White, S. D. M.
        1985, \apjs, 57, 241

\bibitem[Hockney \& Eastwood(1981)]{he81}
        Hochney, R. W., \& Eastwood, J. W. 1981, Computer Simulation Using 
        Particles (New York: McGraw-Hill).

\bibitem[Press et al(1992)]{numrec} Press, W. H., Teukolsky, S. A.,
        Vettering, W. T., \& Flannery, B. P. 1992, Numerical Recipes
        (Cambridge University Press).

\bibitem[Zel'dovich(1970)]{zeldovich70}
        Zel'dovich, Ya. B. 1970, A\&A, 5, 84

\end{thebibliography}
\end{document}